\documentclass[aps,prd,amsfonts,amssymb, amsmath, showpacs, showkeys, a4paper, nofootinbib, superscriptaddress, 11pt, raggedbottom]{revtex4-2}

\usepackage{braket}
\usepackage{bm}
\usepackage{graphicx}
\usepackage{slashed}
\usepackage{subfig}
\usepackage{mathrsfs}
\usepackage{color}
\usepackage{mathtools}
\usepackage[normalem]{ulem}
\usepackage{simplewick}

\DeclareMathOperator*{\sumint}{%
\mathchoice%
  {\ooalign{$\displaystyle\sum$\cr\hidewidth$\displaystyle\int$\hidewidth\cr}}
  {\ooalign{\raisebox{.14\height}{\scalebox{.7}{$\textstyle\sum$}}\cr\hidewidth$\textstyle\int$\hidewidth\cr}}
  {\ooalign{\raisebox{.2\height}{\scalebox{.6}{$\scriptstyle\sum$}}\cr$\scriptstyle\int$\cr}}
  {\ooalign{\raisebox{.2\height}{\scalebox{.6}{$\scriptstyle\sum$}}\cr$\scriptstyle\int$\cr}}
}

\begin{document}

\title{\Large Directly computing Wigner functions for open quantum systems}

\author{Nick Huggett}
\email{huggett@uic.edu}
\affiliation{University of Illinois Chicago,
Chicago, Illinois, USA}

\author{Christian K\"{a}ding}
\email{christian.kaeding@tuwien.ac.at}
\affiliation{Atominstitut, Technische Universit\"at Wien, Stadionallee 2, 1020 Vienna, Austria}


\author{Mario Pitschmann}
\email{mario.pitschmann@tuwien.ac.at}
\affiliation{Atominstitut, Technische Universit\"at Wien, Stadionallee 2, 1020 Vienna, Austria}

\author{James Read}
\email{james.read@philosophy.ox.ac.uk}
\affiliation{Faculty of Philosophy, University of Oxford, Oxford, UK}

\begin{abstract}
The Wigner function is a well-known phase space distribution function with many applications in quantum mechanics. In this article, we consider an open quantum system consisting of a non-relativistic single particle interacting with a general, possibly relativistic environment. For this system, we derive an expression for directly computing the time-dependent Wigner function from its initial values. This result renders time-dependent Wigner functions more applicable without having to make additional approximations that would otherwise be required in order to solve the corresponding equation of motion. As an illustration of our findings, we discuss the example of a non-relativistic single scalar particle interacting via a Yukawa interaction with an environment comprising another type of scalar field that is treated relativistically. 
\end{abstract}


\maketitle



\section{Introduction}

Realistic quantum systems usually interact with one or multiple other systems, their so-called environments. Since the dynamics of environments are often intractable and since we are often interested only in their impact on the quantum system, it is customary to introduce the notion of an open quantum system \cite{Breuer2002}. For this purpose, the environmental degrees of freedom are traced out, leaving an effective description of the system of interest under the influence of the environments. Applications of open quantum systems range from non-relativistic quantum mechanics, see, for example, Refs.~\cite{Carmichael,Gardiner2004,Walls2008,Aolita2015,Goold2016,Werner2016,Huber2020,Keefe:2024cia,Nishimura:2024one,Cavina:2025xnz,Nishimura:2025dam,Coppola:2025qxk,Bose:2025ogr,Bose:2025tbz,Yadav:2025jbi}, to quantum field theory \cite{Calzetta2008,Koksma2010,Koksma2011,Sieberer2016,Marino2016,Baidya2017,Burrage2018,Burrage2019,Banerjee:2020ljo,Nagy2020,Banerjee:2021lqu,Jana2021,Fogedby2022,Kading:2022jjl,Cao:2023syu,Bowen:2024emo,Reyes-Osorio:2024chg,Fahn:2024fgc,Kashiwagi:2024fuy,Salcedo:2024nex,Burrage:2025xac,Farooq:2025pxt,Manshouri:2025nqc}, cosmology \cite{Lombardo1,Lombardo2,Lombardo3,Boyanovsky1,Boyanovsky2,Boyanovsky3,Boyanovsky4,Burgess2015,Hollowood,Shandera:2017qkg,Choudhury:2018rjl,Bohra:2019wxu,Akhtar:2019qdn,Binder2021,Brahma:2021mng,Cao:2022kjn,Brahma2022,Colas:2022hlq,Colas:2022kfu,Colas:2023wxa,Kading:2023mdk,Bhattacharyya:2024duw,Colas:2024xjy,Burgess:2024eng,Salcedo:2024smn,Colas:2024lse,Colas:2024ysu,Brahma:2024yor,Kading:2024jqe,Brahma:2024ycc,Burgess:2024heo,Lau:2024mqm,Takeda:2025cye,Nandi:2025qyj,Cho:2025vjo,Kranas:2025jgm,Salcedo:2025ezu,Cespedes:2025zqp,Burgess:2025dwm}, black hole physics \cite{Yu2008,Lombardo2012,Jana2020,Agarwal2020,Kaplanek2020,Burgess:2021luo,Kaplanek2021} or heavy ion physics \cite{Brambilla1,Brambilla2,Yao2018,Yao2020,Akamatsu2020,DeJong2020,Yao2021,Brambilla2021,Griend2021,Yao2022}.

While reduced density matrices are often the tools of choice to mathematically describe an open quantum system, Wigner functions \cite{Wigner} are closely related alternatives that allow for discussions of phase spaces in quantum mechanics and are, therefore, particularly useful to illustrate open quantum dynamical effects, for example, transitions from quantum to classical states. The time evolution of Wigner functions can be expressed in terms of so-called Wigner flows \cite{PhysRevLett.87.223202,Bauke} that reveal otherwise hidden features of quantum dynamics \cite{PhysRevLett.110.030401}; see Ref.~\cite{PhysRevA.100.012124} for a discussion of Wigner flows for open systems. Wigner functions have found a variety of applications, see, e.g., Refs.~\cite{mandel1965coherence,barker1983quasi,frensley1987wigner,kim1990canonical,kohen1997phase,SudaBook,coffey2007wigner,vacchini2007relaxation,croitoru2008quantum,querlioz2009wigner,alonso2011wigner,veitch2013efficient,Giese:2014laa,ferry2015phase,xu2015quantum,ren2016evolution,wigger2016quantum,papendell2017quantum,Kohlfurst:2021skr,Suda:2021pit,simoncelli2022wigner,hahn2022photon,Kohlfurst:2022edl,Amin:2025sdc,Mandal:2025ofl,Kiamari:2025ios,Azuma:2025xgf,Kar:2025qvj}, and have been thoroughly reviewed in Refs.~\cite{HILLERY1984121,LEE1995147,10.1119/1.2957889,FerryBook,10.1063/1.5046663,10.1002/qute.202100016,encyclopedia5030118}. 

In this article, we consider an open quantum system consisting of a non-relativistic single particle interacting with a general environment, which may be relativistic. A non-relativistic description of the system under consideration enables a far simpler investigation of certain quantum theoretical questions, such as, e.g.,\ the emergence of classicality, than a relativistic one. For this system, inspired by the methods for directly computing density matrices that were introduced in Refs.~\cite{Kading:2022jjl,Kading:2022hhc} and which already had phenomenological applications in Refs.~\cite{Kading:2023mdk,Kading:2024jqe,Kading:2025azy}, we present a new approach for directly computing time-dependent Wigner functions from their values at an initial time without having to solve an equation of motion. Besides the system being non-relativistic, we merely make the usual assumption that the system and its environment(s) were separated at the initial time and that the considered interaction is sufficiently weak to justify a perturbative treatment. This approach circumvents having to solve analytically intricate equations of motion for Wigner functions. As a consequence, we expect it to open up opportunities for novel applications of Wigner functions or for improvements of existing ones.

The article is organized as follows. In Sec.~\ref{sec:Wigner}, we derive an expression for directly computing Wigner functions for a general non-relativistic single particle in interaction with a general environment. Subsequently, in Sec.~\ref{sec:Yukawa}, we discuss the concrete example of two scalar fields interacting via a Yukawa coupling in order to illustrate our findings. Finally, we draw our conclusions in Sec.~\ref{sec:Conclusion}.


\section{Wigner function}
\label{sec:Wigner}

In this section, we derive a general expression for directly computing the time-dependent Wigner function for a non-relativistic single particle acting as an open quantum system after tracing out the degrees of freedom of an environment with which the particle is interacting. At first, we consider a general quantum system $\mathcal{S}$ and its environment $\mathcal{E}$. Before system and environment start to interact at time $t=0$, the total density operator is the direct product of both subsystem density operators, i.e., for times $t\leq0$:
\begin{eqnarray}
\hat\rho(t) 
&=& \hat\rho_\mathcal{S}(t)\otimes\hat\rho_\mathcal{E}(t)~.
\end{eqnarray}
In the interaction representation, the density operator for times $t\geq0$ is given by \cite{Kading:2025cwg}
\begin{eqnarray}
\hat\rho(t) &=& \mathbf{T}\!\left(e^{ -\mathrm{i}\int_0^tdt'\,\hat H^{(I)}(t')}\right)\hat\rho(0)\,\mathbf{\bar T}\!\left(e^{\mathrm{i}\int_0^tdt'\,\hat H^{(I)}(t')}\right) ~,
\end{eqnarray}
where $\hat H^{(I)}$ is the Hamiltonian describing all interactions (including self-interactions) of the total system combining system and environment, and $\mathbf{T}$ and $\mathbf{\bar T}$ denote time-ordering and anti-time-ordering, respectively. From this, we obtain the reduced density operator of the system $\mathcal{S}$ by tracing out the environmental degrees of freedom:
\begin{eqnarray}
\hat\rho_\mathcal{S}(t) 
&=& 
\text{Tr}_\mathcal{E}\left[ \hat\rho(t) \right]
\nonumber
\\
&=& 
\sumint_m \big(\mathbb{I}\otimes\langle m|\big)
\hat\rho(t)
\big(\mathbb{I}\otimes|m\rangle\big)
\nonumber
\\
&=& 
\sumint_m \big(\mathbb{I}\otimes\langle m|\big)
\mathbf{T}\!\left(e^{ -\mathrm{i}\int_0^tdt'\,\hat H^{(I)}(t')}\right)\hat\rho_\mathcal{S}(0)\otimes\hat\rho_\mathcal{E}(0)\,\mathbf{\bar T}\!\left(e^{\mathrm{i}\int_0^tdt'\,\hat H^{(I)}(t')}\right)
\big(\mathbb{I}\otimes|m\rangle\big)
~,
\end{eqnarray}
where the trace does not depend on the particular choice for the basis $\ket{m}$ as long as the environmental Hilbert space is spanned by it. Since this basis can, in general, either be discrete or continuous, $\sumint$ denotes sums as well as integrations. The time-dependent Wigner function of the system is a Fourier transform of the reduced density matrix \cite{Wigner} (see also Ref.~\cite{FerryBook}):
\begin{eqnarray}
\label{eq:Wigner}
    W_\mathcal{S}(\vec{x};\vec{p};t) 
&=&
\frac{1}{\pi^3} \int d^3z \rho_\mathcal{S}(\vec{x}-\vec{z},\vec{x}+\vec{z};t) e^{2\mathrm{i}\vec{p}\vec{z}}~.
\end{eqnarray}

In order to derive an expression that will allow us to directly compute the Wigner function at time $t$ from its initial values at time $0$, we will from now on only focus on the system $\mathcal{S}$ being a non-relativistic particle with mass $m_\mathcal{S}$. In this case, the reduced single-particle density matrix elements in position space are given by
\begin{eqnarray}
\rho_\mathcal{S}(\vec{x},\vec{y};t) 
&=&
\bra{0} \hat{\Psi}(\vec{x},t) \hat{\rho}_\mathcal{S}(t) \hat{\Psi}^\dagger(\vec{y},t) \ket{0}
\nonumber
\\
&=& 
\sumint_m \big(\bra{0} \hat{\Psi}(\vec{x},t)\otimes\langle m|\big)
\mathbf{T}\!\left(e^{ -\mathrm{i}\int_0^tdt'\,\hat H^{(I)}(t')}\right)\hat\rho_\mathcal{S}(0)\otimes\hat\rho_\mathcal{E}(0)
\nonumber
\\
&&
~~~~~~~~~~~~~~~~~~~~~~~~~~~
\times
\mathbf{\bar T}\!\left(e^{\mathrm{i}\int_0^tdt'\,\hat H^{(I)}(t')}\right)
\big(\hat{\Psi}^\dagger(\vec{y},t) \ket{0}\otimes|m\rangle\big)
~,
\end{eqnarray}
where $\hat{\Psi}$ is the second quantized field operator for a non-relativistic particle, which creates a particle located at $\vec{x}$, i.e.,\ $|\vec{x}\rangle=\hat\Psi^\dagger(\vec{x})|0\rangle$. Since the one-particle unit operator is given by
\begin{eqnarray}
    \mathbb{I}_1 
    &:=& 
    \int d^3z \hat{\Psi}^\dagger(\vec{z},0) \ket{0}\bra{0} \hat{\Psi}(\vec{z},0)~,
\end{eqnarray}
we obtain
\begin{eqnarray}
\rho_\mathcal{S}(\vec{x},\vec{y};t) 
&=& 
\sumint_m \big(\bra{0} \hat{\Psi}(\vec{x},t)\otimes\langle m|\big)
\mathbf{T}\!\left(e^{ -\mathrm{i}\int_0^tdt'\,\hat H^{(I)}(t')}\right)\mathbb{I}_1\hat\rho_\mathcal{S}(0)\mathbb{I}_1
\nonumber
\\
&&
\otimes\hat\rho_\mathcal{E}(0)
\mathbf{\bar T}\!\left(e^{\mathrm{i}\int_0^tdt'\,\hat H^{(I)}(t')}\right)
\big(\hat{\Psi}^\dagger(\vec{y},t) \ket{0}\otimes|m\rangle\big)
\nonumber
\\
&=& 
\sumint_m \big(\bra{0} \hat{\Psi}(\vec{x},t)\otimes\langle m|\big)
\mathbf{T}\!\left(e^{ -\mathrm{i}\int_0^tdt'\,\hat H^{(I)}(t')}\right)\int d^3z \hat{\Psi}^\dagger(\vec{z},0) \ket{0}\bra{0} \hat{\Psi}(\vec{z},0)\hat\rho_\mathcal{S}(0)
\nonumber
\\
&&
\times
\int d^3z' \hat{\Psi}^\dagger(\vec{z}^{\,'},0) \ket{0}\bra{0} \hat{\Psi}(\vec{z}^{\,'},0)\otimes\hat\rho_\mathcal{E}(0)
\mathbf{\bar T}\!\left(e^{\mathrm{i}\int_0^tdt'\,\hat H^{(I)}(t')}\right)
\big(\hat{\Psi}^\dagger(\vec{y},t) \ket{0}\otimes|m\rangle\big)
\nonumber
\\
&=&
\int d^3z d^3z'
\rho_\mathcal{S}(\vec{z},\vec{z}^{\,'};0)
\sumint_m 
\big(\bra{0} \hat{\Psi}(\vec{x},t)\otimes\langle m|\big)
\mathbf{T}\!\left(e^{ -\mathrm{i}\int_0^tdt'\,\hat H^{(I)}(t')}\right)
\nonumber
\\
&&
\times
\left(  
\hat{\Psi}^\dagger(\vec{z},0)
\ket{0}\bra{0} 
\hat{\Psi}(\vec{z}^{\,'},0)
\otimes
\hat{\rho}_\mathcal{E}(0)
\right)
\mathbf{\bar T}\!\left(e^{\mathrm{i}\int_0^tdt'\,\hat H^{(I)}(t')}\right)
\big(\hat{\Psi}^\dagger(\vec{y},t) \ket{0}\otimes|m\rangle\big)
~,
\end{eqnarray}
where we have used that 
\begin{eqnarray}
   \rho_\mathcal{S}(\vec{z},\vec{z}^{\,'};0) 
&=& 
\bra{0}
\hat{\Psi}(\vec{z},0)\hat\rho_\mathcal{S}(0)
\hat{\Psi}^\dagger(\vec{z}^{\,'},0) \ket{0}~.
\end{eqnarray}
Next, we split the Hamiltonian:
\begin{eqnarray}
    \hat{H}^{(I)}(t) &=& \hat{H}^{(I)}_\mathcal{S}(t) \otimes \mathbb{I} + \mathbb{I} \otimes \hat{H}^{(I)}_\mathcal{E}(t) + \hat{H}^{(I)}_{\mathcal{S}\mathcal{E}}(t)~,
\end{eqnarray}
where $\hat{H}^{(I)}_\mathcal{S}$ and $\hat{H}^{(I)}_\mathcal{E}$ describe the self-interactions of system and environment, respectively, and $\hat{H}^{(I)}_{\mathcal{S}\mathcal{E}}$ denotes the Hamiltonian that captures the interaction between system and environment. 
Consequently, we find
\begin{eqnarray}
\label{eq:DensMatrFinal}
\rho_\mathcal{S}(\vec{x},\vec{y};t) 
&=&
\int d^3z d^3z'
\rho_\mathcal{S}(\vec{z},\vec{z}^{\,'};0)
\sumint_m 
\big(\bra{0} \hat{\Psi}(\vec{x},t)\otimes\langle m|\big)
\mathbf{T}\!\left(e^{ -\mathrm{i}\int_0^tdt'\,\big[\hat{H}^{(I)}_\mathcal{S}(t') \otimes \mathbb{I} + \mathbb{I} \otimes \hat{H}^{(I)}_\mathcal{E}(t') + \hat{H}^{(I)}_{\mathcal{S}\mathcal{E}}(t')\big]}\right)
\nonumber
\\
&&
\times
\left(  
\hat{\Psi}^\dagger(\vec{z},0)
\ket{0}\bra{0} 
\hat{\Psi}(\vec{z}^{\,'},0)
\otimes
\hat{\rho}_\mathcal{E}(0)
\right)
\mathbf{\bar T}\!\left(e^{\mathrm{i}\int_0^tdt'\,\big[\hat{H}^{(I)}_\mathcal{S}(t') \otimes \mathbb{I} + \mathbb{I} \otimes \hat{H}^{(I)}_\mathcal{E}(t') + \hat{H}^{(I)}_{\mathcal{S}\mathcal{E}}(t')\big]}\right)
\nonumber
\\
&&
\times
\big(\hat{\Psi}^\dagger(\vec{y},t) \ket{0}\otimes|m\rangle\big)
~.
\end{eqnarray}

We could now simply substitute Eq.~(\ref{eq:DensMatrFinal}) into Eq.~(\ref{eq:Wigner}) in order to find the Wigner function for the single particle. However, this would give us the Wigner function in terms of the particle’s initial density matrix. Since we are instead interested in computing the Wigner function in terms of its initial values, we must find the inverse of Eq.~(\ref{eq:Wigner}). Multiplying Eq.~(\ref{eq:Wigner}) by $e^{-\mathrm{i}\vec{p}(\vec{y}-\vec{x})}$ and integrating over $\vec{p}$ gives\footnote{Note that, when discussing systems in finite volumes, the integration has to be replaced by a sum over $\vec{p}$.}
\begin{eqnarray}
    \rho_\mathcal{S}(\vec{x},\vec{y};t)
&=&
\int d^3p \,W_\mathcal{S}\left( \frac{\vec{x}+\vec{y}}{2};\vec{p};t \right) e^{-\mathrm{i}\vec{p}(\vec{y}-\vec{x})}~.
\end{eqnarray}
This expression allows us to replace the initial density matrix elements by initial values of the Wigner function. Consequently, we arrive at
\begin{eqnarray}
\label{eq:WignerFinal}
        W_\mathcal{S}(\vec{x};\vec{p};t) 
&=&
\frac{1}{(2\pi)^3}\int d^3y d^3z d^3z' \int d^3q e^{ \mathrm{i} \vec{p}\vec{y} } e^{-\mathrm{i} \vec{q} (\vec{z}^{\,'}-\vec{z})}
W_\mathcal{S}\left( \frac{\vec{z}+\vec{z}^{\,'}}{2};\vec{q} ;0 \right)
\nonumber
\\
&&
\times
\sumint_m 
\left(\bra{0} \hat{\Psi} \left(\vec{x}-\frac{\vec{y}}{2},t\right)\otimes\bra{m}\right)
\mathbf{T}\!\left(e^{ -\mathrm{i}\int_0^tdt'\,\big[\hat{H}^{(I)}_\mathcal{S}(t') \otimes \mathbb{I} + \mathbb{I} \otimes \hat{H}^{(I)}_\mathcal{E}(t') + \hat{H}^{(I)}_{\mathcal{S}\mathcal{E}}(t')\big]}\right)
\nonumber
\\
&&
\times
\left(  
\hat{\Psi}^\dagger(\vec{z},0)
\ket{0}\bra{0} 
\hat{\Psi}(\vec{z}^{\,'},0)
\otimes
\hat{\rho}_\mathcal{E}(0)
\right)
\mathbf{\bar T}\!\left(e^{\mathrm{i}\int_0^tdt'\,\big[\hat{H}^{(I)}_\mathcal{S}(t') \otimes \mathbb{I} + \mathbb{I} \otimes \hat{H}^{(I)}_\mathcal{E}(t') + \hat{H}^{(I)}_{\mathcal{S}\mathcal{E}}(t')\big]}\right)\nonumber
\\
&&
\times
\left(\hat{\Psi}^\dagger\left(\vec{x}+\frac{\vec{y}}{2},t\right) \ket{0}\otimes\ket{m}\right)~. 
\end{eqnarray}
We would like to note that this expression allows to directly compute the Wigner function at any later time from given initial data. This is significantly more tractable than the standard approach by solving Moyal's evolution equation~\cite{Moyal:1949sk,Marzlin:2015lyw}, which is often analytically intricate or impossible. Therefore, we expect our approach to broaden the range of applications of time-dependent Wigner functions.


\section{Example: Yukawa interaction}
\label{sec:Yukawa}

\begin{figure}[htbp]
\centering
\subfloat[][]{\includegraphics[scale=0.5]{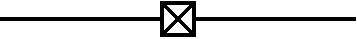}}
\qquad
\subfloat[][]{\includegraphics[scale=0.5]{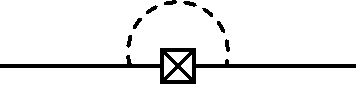}}

\subfloat[][]{\includegraphics[scale=0.5]{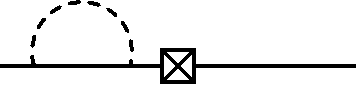}}
\qquad
\subfloat[][]{\includegraphics[scale=0.5]{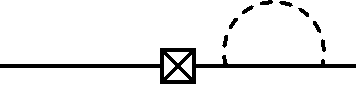}}

\caption{\label{Fig:Diags} Diagrammatic representation of the terms contributing to $W_\mathcal{S}(\vec{x};\vec{p};t)$ as given in Eq.~(\ref{eq:WignerExFinal}); a crossed box represents the Wigner function at the initial time $0$, a solid line stands for a propagator of the non-relativistic particle, and a dashed line is a $\phi$-propagator. (a): free evolution of the non-relativistic particle corresponding to the zeroth order term in Eq.~(\ref{eq:WignerExFinal}); (b), (c) and (d) represent the first, second and third term at order $g^2$, respectively.}
\end{figure}

We now illustrate the result from the previous section by discussing a concrete example. For this, we consider a non-relativistic single scalar particle as the system and a relativistic real scalar field $\phi$ with mass $m_\mathcal{E}$ as the environment. Within an infinite spatial volume, they interact via a Yukawa coupling described by the Hamiltonian
\begin{eqnarray}
\label{eq:YukawaHam}
    \hat{H}^{(I)}_{\mathcal{S}\mathcal{E}}(t)
&=&
- g\int d^3x \hat{\Psi}(x)\hat{\Psi}^\dagger(x)\otimes \hat{\phi}(x)~,
\end{eqnarray}
where $g$ is the coupling constant. For simplicity, we assume that neither the system nor the environment is self-interacting. In order to evaluate Eq.~(\ref{eq:WignerFinal}) for the chosen interaction, we must employ perturbation theory and will restrict our discussion to terms up to second order in $g$: 
\begin{eqnarray}
        W_\mathcal{S}(\vec{x};\vec{p};t) 
&=&
\frac{1}{(2\pi)^3}\int d^3y d^3z d^3z' \int d^3q e^{ \mathrm{i} \vec{p}\vec{y} } e^{-\mathrm{i} \vec{q} (\vec{z}^{\,'}-\vec{z})}
W_\mathcal{S}\left( \frac{\vec{z}+\vec{z}^{\,'}}{2};\vec{q} ;0 \right)
\nonumber
\\
&&
\times
\sumint_m
\bigg[\bra{0} \hat{\Psi} \left(\vec{x}-\frac{\vec{y}}{2},t\right)\otimes\langle m|\bigg]
\bigg[ \mathbb{I}\otimes\mathbb{I} + \mathrm{i}g \int d^4x' 
\hat{\Psi}(x')\hat{\Psi}^\dagger(x')\otimes \hat{\phi}(x')
\nonumber
\\
&&
~~~
-
\frac{g^2}{2} 
\mathbf{T} 
\int d^4x'  d^4y' 
\hat{\Psi}(x')\hat{\Psi}^\dagger(x')\hat{\Psi}(y')\hat{\Psi}^\dagger(y')\otimes \hat{\phi}(x') \hat{\phi}(y')
\bigg]
\nonumber
\\
&&
~
\times
\left(  
\hat{\Psi}^\dagger(\vec{z},0)
\ket{0}\bra{0} 
\hat{\Psi}(\vec{z}^{\,'},0)
\otimes
\hat{\rho}_\mathcal{E}(0)
\right)
\bigg[ \mathbb{I}\otimes\mathbb{I} - \mathrm{i}g \int d^4x' 
\hat{\Psi}(x')\hat{\Psi}^\dagger(x')\otimes \hat{\phi}(x')
\nonumber
\\
&&
-
\frac{g^2}{2} 
\mathbf{\bar T}
\int d^4x'  d^4y' 
\hat{\Psi}(x')\hat{\Psi}^\dagger(x')\hat{\Psi}(y')\hat{\Psi}^\dagger(y')\otimes \hat{\phi}(x') \hat{\phi}(y')
\bigg]
\bigg[\hat{\Psi}^\dagger\left(\vec{x}+\frac{\vec{y}}{2},t\right) \ket{0}\otimes|m\rangle\bigg]
\nonumber
\\
&&
+ \mathcal{O}(g^3)
~. 
\end{eqnarray}
We can evaluate the trace over the environmental degrees of freedom by using the relativistic scalar field Feynman, Dyson and Wightman propagators:
\begin{eqnarray}
\Delta^\text{F}_{xy} &:=&
   \sumint_m\langle m| \mathbf{T}\hat{\phi}(x) \hat{\phi}(y) \hat{\rho}_\mathcal{E}(0)|m \rangle~,
\\
\Delta^\text{D}_{xy} &:=&
   \sumint_m\langle m| \mathbf{\bar T}\hat{\phi}(x) \hat{\phi}(y) \hat{\rho}_\mathcal{E}(0) |m \rangle~,
\\
\Delta^{<}_{xy}
&:=&
\sumint_m\langle m| \hat{\phi}(y)\hat{\phi}(x) \hat{\rho}_\mathcal{E}(0) |m \rangle~,
\end{eqnarray}
so that we are left with (after dropping the $\mathcal{O}(g^3)$ for notational convenience):
\begin{eqnarray}
\label{eq:WigBsp1}
        W_\mathcal{S}(\vec{x};\vec{p};t) 
&\approx&
\frac{1}{(2\pi)^3}\int d^3y d^3z d^3z' \int d^3q e^{ \mathrm{i} \vec{p}\vec{y} } e^{-\mathrm{i} \vec{q} (\vec{z}^{\,'}-\vec{z})}
W_\mathcal{S}\left( \frac{\vec{z}+\vec{z}^{\,'}}{2};\vec{q} ;0 \right)
\nonumber
\\
&&
\times
\bra{0} \hat{\Psi} \left(\vec{x}-\frac{\vec{y}}{2},t\right)
\bigg\{
\hat{\Psi}^\dagger(\vec{z},0)
\ket{0}\bra{0} 
\hat{\Psi}(\vec{z}^{\,'},0)
\nonumber
\\
&&
~~
+
g^2 \int d^4x' d^4y' 
\bigg[
\Delta^{<}_{x'y'}
\hat{\Psi}(x')\hat{\Psi}^\dagger(x')
\hat{\Psi}^\dagger(\vec{z},0)
\ket{0}\bra{0} 
\hat{\Psi}(\vec{z}^{\,'},0)
\hat{\Psi}(y')\hat{\Psi}^\dagger(y')
\nonumber
\\
&&
~~~
-
\frac{1}{2} \Delta^\text{F}_{x'y'} 
\mathbf{T} 
\hat{\Psi}(x')\hat{\Psi}^\dagger(x')\hat{\Psi}(y')\hat{\Psi}^\dagger(y')
\hat{\Psi}^\dagger(\vec{z},0)
\ket{0}\bra{0} 
\hat{\Psi}(\vec{z}^{\,'},0)
\nonumber
\\
&&
~~~
-
\frac{1}{2} \Delta^\text{D}_{x'y'} 
\hat{\Psi}^\dagger(\vec{z},0)
\ket{0}\bra{0} 
\hat{\Psi}(\vec{z}^{\,'},0)
\mathbf{\bar T} 
\hat{\Psi}(x')\hat{\Psi}^\dagger(x')\hat{\Psi}(y')\hat{\Psi}^\dagger(y')
\bigg] 
\bigg\}
\nonumber
\\
&&
~
\times
\hat{\Psi}^\dagger\left(\vec{x}+\frac{\vec{y}}{2},t\right) \ket{0}
~. 
\end{eqnarray}
Next, we introduce non-relativistic Feynman and Dyson propagators for the system particle:
\begin{eqnarray}
\label{eq:Green++}
G^\text{F}(x;y)
&:=& 
\langle0| \mathbf{T} \hat{\Psi}(x) \hat{\Psi}^\dagger(y)|0 \rangle
=
-\int\frac{d\omega}{2\pi i}\int\frac{d^3p}{(2\pi)^3}\,\frac{e^{-i\omega(x^0 - y^0)}\,e^{i\vec{p}\cdot(\vec{x} - \vec{y})}}{\omega - \frac{\vec{p}^{\,2}}{2m_\mathcal{S}} + i\varepsilon}~,
\\
G^\text{D}(x;y)
&:=& 
\langle0| \mathbf{\bar T} \hat{\Psi}(x) \hat{\Psi}^\dagger(y)|0 \rangle
=
\int\frac{d\omega}{2\pi i}\int\frac{d^3p}{(2\pi)^3}\,\frac{e^{-i\omega(x^0 - y^0)}\,e^{i\vec{p}\cdot(\vec{x} - \vec{y})}}{\omega - \frac{\vec{p}^{\,2}}{2m_\mathcal{S}} - i\varepsilon}~.
\end{eqnarray}
Note that, in the non-relativistic case, the following holds:
\begin{eqnarray}
    \langle \hat{\Psi}(x) \hat{\Psi}^\dagger(y) \rangle
&=& 
\Theta(x^0-y^0)G^\text{F}(x;y) + \Theta(y^0-x^0)G^\text{D}(x;y)
~.
\end{eqnarray}
Consequently, Eq.~(\ref{eq:WigBsp1}) becomes
\begin{eqnarray}
\label{eq:WignerExFinal}
        &&W_\mathcal{S}(\vec{x};\vec{p};t) 
    \nonumber
\\
&&
\approx
\frac{1}{(2\pi)^3}\int d^3y d^3z d^3z' \int d^3q e^{ \mathrm{i} \vec{p}\vec{y} } e^{-\mathrm{i} \vec{q} (\vec{z}^{\,'}-\vec{z})}
W_\mathcal{S}\left( \frac{\vec{z}+\vec{z}^{\,'}}{2};\vec{q} ;0 \right)
\nonumber
\\
&&
\times
\bigg\{
G^\text{F}\left( \vec{x} - \frac{\vec{y}}{2},t; \vec{z},0 \right)
G^\text{D}\left(\vec{z}{\,'},0;  \vec{x} + \frac{\vec{y}}{2},t \right)
\nonumber
\\
&&
~~
+
g^2 \int d^4x' d^4y' 
\bigg[
\Delta^{<}_{x'y'}
G^\text{F}\left( \vec{x} - \frac{\vec{y}}{2},t; x' \right)
G^\text{F}\left(  x'; \vec{z},0 \right)
G^\text{D}\left(  \vec{z}^{\,'},0; y' \right)
G^\text{D}\left( y'; \vec{x} + \frac{\vec{y}}{2},t \right)
\nonumber
\\
&&
~~~~~~~~~~~~~~~~~~~~~
-
\Delta^\text{F}_{x'y'} 
G^\text{F}\left( \vec{x} - \frac{\vec{y}}{2},t; x' \right)
G^\text{F}\left(  y'; x' \right)
G^\text{F}\left(  y'; \vec{z},0 \right)
G^\text{D}\left( \vec{z}^{\,'},0; \vec{x} + \frac{\vec{y}}{2},t \right)
\nonumber
\\
&&
~~~~~~~~~~~~~~~~~~~~~
-
\Delta^\text{D}_{x'y'} 
G^\text{F}\left( \vec{x} - \frac{\vec{y}}{2},t; \vec{z},0 \right)
G^\text{D}\left(  x'; y' \right)
G^\text{D}\left(  \vec{z}^{\,'},0; x' \right)
G^\text{D}\left( y'; \vec{x} + \frac{\vec{y}}{2},t \right)
\bigg] 
\bigg\}
~,~~~
\end{eqnarray}
where we have dropped all terms corresponding to disconnected diagrams. Figure \ref{Fig:Diags} depicts all diagrams that we have taken into account.

At this point, we end our discussion of the example since, in order to proceed significantly further, we would have to choose an initial Wigner function---unfortunately, however, only a few particular choices would render the computation analytically tractable. While we could use Eq.~(\ref{eq:Wigner}) in order to replace the initial Wigner function by an initial density matrix, this would also thwart our approach of employing a closed expression for Wigner functions and is therefore not pursued here. Nevertheless, already in its current form, Eq.~(\ref{eq:WignerExFinal}) demonstrates that the method developed in the present article indeed provides a directly computable expression for a time-dependent Wigner function from its initial values.

\section{Conclusion}
\label{sec:Conclusion}

Open quantum systems are effective descriptions of systems under the influence of environments whose degrees of freedom have been traced out mathematically. Powerful tools for describing the open dynamics of such systems are Wigner functions. In this article, we have introduced a new method for directly computing Wigner functions from their initial values. For this purpose, we have restricted our discussion to systems comprising a non-relativistic single particle interacting with general, potentially relativistic environments. In addition, we have assumed that the system and its environments are separated at the initial time and that the interaction is weak enough to allow us to use perturbation theory. Since this novel method enables us to circumvent equations of motion for Wigner functions that can be analytically intricate to solve, it is expected to create avenues to additional applications of Wigner functions or to refinements of already existing ones. In order to demonstrate the utility of the developed approach, we have discussed a simple example of a non-relativistic single scalar particle interacting via a Yukawa coupling with an environment comprising a relativistic scalar field of another type. Deeper analyses of more realistic quantum systems and finite-time effects using the presented method will be the subject of future works. An important example would be decoherence. Decoherence is, of course, also related intimately to the emergence of classicality in quantum systems (see Refs. \cite{Zurek1981, Zurek1982, Zurek1991, Zurek1993, AnglinZurek1995, DalvitDziarmagaZurek2005, Zurek2003}); our work also offers one further outlook on this phenomenon. Going forward, it would be a valuable exercise to compute numerically the time evolution of an initial non-classical Wigner function for an open system, and witness the emergence of a classical probability distribution in phase space---one would expect the results of these computations to be in accord with the discussions of the evolution of Wigner functions presented in e.g.\ Ref.\ \cite{Zurek1991}. We leave further exploration of this to future work.


\begin{acknowledgments}
The authors are very grateful to Daniel Grimmer for useful discussions and comments on the manuscript. This research was funded in whole or in part by the Austrian Science Fund (FWF) 
[10.55776/PAT7599423] and [10.55776/PAT8564023]. For open access purposes, the authors have applied a CC BY public copyright license to any author accepted manuscript version arising from this submission.
\end{acknowledgments}



\bibliography{Bib}
\bibliographystyle{JHEP}

\end{document}